\documentstyle[12pt,aasms4]{article}

\slugcomment{Accepted for publication in ApJ, Part I}

\lefthead{Gibson}
\righthead{Photo-chemical Evolution Code Comparison}

\begin{document}

\title{\hspace{-5.0mm}A Comparison of Three Elliptical Galaxy \\
    Photo-chemical Evolution Codes}

\author{Brad K. Gibson}
\affil{Mount Stromlo \& Siding Spring Observatories, \\
    Australian National University, \\
    Weston Creek P.O., Weston, ACT, 2611, Australia}

\vskip3.0truemm

\def\spose#1{\hbox to 0pt{#1\hss}}
\def\simlt{\mathrel{\spose{\lower 3pt\hbox{$\mathchar"218$}}
     \raise 2.0pt\hbox{$\mathchar"13C$}}}
\def\simgt{\mathrel{\spose{\lower 3pt\hbox{$\mathchar"218$}}
     \raise 2.0pt\hbox{$\mathchar"13E$}}}
\def\eg{{\rm e.g. }}
\def\ie{{\rm i.e. }}
\def\etal{{\rm et~al. }}

\begin{abstract}
Working within the classic supernovae-driven wind framework for elliptical
galaxy evolution, we perform a systematic investigation into the discrepancies
between the predictions of three contemporary codes -- Arimoto \& Yoshii (1987,
A\&A, 173, 23), Bressan \etal (1994, ApJS, 94, 63), and Gibson
(1996a, MNRAS, 278, 829; 1996b, MNRAS, submitted).  By being primarily concerned
with reproducing the present-day colour-metallicity-luminosity (CML) relations
amongst ellipticals, the approaches
taken in the theoretical modelling have managed 
to obscure many of the hidden differences between the codes.
Targeting the timescale for the onset of the initial galactic wind $t_{\rm
GW}$ as a primary ``difference''
indicator, we demonstrate exactly how and why each code 
is able to claim successful reproduction of the CML relations, despite
possessing apparently incompatible input ingredients.
\end{abstract}


\section{Introduction}
\label{introduction}

In one of his early seminal papers, Larson (1974) postulated that the 
mass-metallicity relation observed in the present-day
elliptical galaxy population (\eg \cite{F77}) was a natural consequence of 
supernovae (SNe)-driven galactic winds.  With each SN contributing
$\sim 10^{50}$ erg of thermal energy to a galaxy's interstellar medium (ISM),
Larson demonstrated that standard star formation rate scenarios would
inevitably result in the expulsion of any remaining gas once its accumulated
thermal energy exceeded that of its gravitational binding energy.
The bulk of subsequent star formation would then be suppressed (at this time
$t_{\rm GW}$), thereby ``freezing'' in the chemical imprints which 
would be observable today in the stellar populations.
Because of a larger binding energy per
unit mass, $t_{\rm GW}$ would occur later in more massive systems, thereby
allowing the enrichment process to progress further than in more massive
systems, in rough accordance with the observed mass-metallicity relation.

The elegant simplicity of Larson's (1974) model was immediately recognized and
provided the basis for a generation of elliptical galaxy
SNe-driven wind models.  The
primary motivation behind the majority of subsequent studies has been to better
understand the chemical (\eg \cite{MT87}; \cite{MT94}; \cite{M94})
and dynamical (\eg \cite{S79b}; \cite{DS86}; \cite{CDPR91})
evolution of these systems.  Related to this has been the
particular attention paid to the role
played by cluster ellipticals in polluting the intracluster medium of galaxy
clusters with metals via large scale super-winds 
(\eg \cite{LD75}; \cite{MV88}; \cite{NC95}; \cite{GM96}).

Transcending from the mass-metallicity plane, to a more
convenient observational colour-metallicity-luminosity (CML) space, requires
the parallel computation of a system's photometric evolution.  This is an
important step to take, as the photometric properties of galaxies should
be a primary constraint in any galactic evolution code.  Until recently,
though, this has been a difficult prospect, owing to the dearth of available
self-consistent, metallicity-sensitive stellar evolution tracks and photometric
calibrations.  This has been alleviated somewhat with the release of the Kurucz
(1993) grid of model atmospheres.  Their subsequent use, in particular
by Worthey (1994) and Bertelli \etal (1994), in constructing extensive grids
of isochrones has made the coupled photo-chemical evolution of ellipticals,
within the SNe-driven wind framework, a realistic aim.

Indeed, during the past few years, several groups have begun investigations in
just this direction -- Bressan \etal (1994) (BCF94, hereafter) 
and Gibson (1996b)\footnote{
Preliminary results can be seen in Gibson
(1994a,b;1995;1996a) and Gibson \& Matteucci (1996).} (G96, hereafter)
are the first of
the new generation of SNe-driven wind codes to take advantage of the coupled
photometric and chemical evolution possibilities.
A precursor, in the same vein as these two more recent codes, belongs to
Arimoto \& Yoshii (1987) (AY87, hereafter).  
Their efforts in piecing together, from
innumerable sources, a grid of stellar evolution tracks and photo-chemical
calibrations, was nothing less than heroic.

In the course of testing our own code, it has become apparent that for all the
similarities in their goals (\ie replicating the present-day CML relations,
primarily) and approaches (\ie classic SNe-driven wind model of \cite{L74}),
there are more than a few subtle, and ``hidden'', differences in the above
three packages which have not been fully appreciated.
In this paper, 
we attempt to bring to light, in as straightforward a manner as
possible, some of these hidden differences.  Considering the proliferation of
such models, we feel it imperative (and long overdue) that such a presentation
be undertaken.

In Section \ref{g96}, we briefly describe a simple template model with which to
work, drawing attention to each of the intrinsic input ingredients (\eg
supernova remnant (SNR) ISM thermal energy deposition efficiency, 
nucleosynthetic yields, initial mass function (IMF)).  The template chosen is
from G96, and successfully reproduces the present-day photo-chemical
properties.  We then systematically explore the influence of varying each of
the primary input ingredients, in accordance with those chosen by AY87 and
BCF94.  This is done in Sections \ref{ay87} and
\ref{bcf94}, respectively.  This is an important exercise which has never been
carried out before, and serves to illustrate exactly how the different groups
were able to piece together seemingly incompatible ingredients and yet,
apparently, recover the proper present-day observable photo-chemical properties.
A summary is provided in Section \ref{summary}.

\section{Comparison With Recent Work}
\label{comparison}

Let us re-iterate once again that the general framework adopted by the three
groups -- AY87, BCF94, and G96 --
is quite similar.  Specifically, elliptical proto-galaxies are taken to be
initially homogeneous gas spheres, and the chemical evolution proceeds
according to the simple closed-box model.  Each avoids the
instantaneous recycling approximation, but subscribes to the instantaneous
mixing approximation.  The global thermal energy budget of the ISM is assumed to
follow the basic prescription outlined in Saito (1979b) and global ejection of
said ISM is taken to occur at $t_{\rm GW}$ (\ie the point at which the thermal
energy exceeds the gravitational binding energy).

Each of the groups claim reproducibility of the fundamental CML relations for
ellipticals.  This is a most interesting result given the very
different assumptions regarding star formation rate efficiencies, dark matter
distributions, SNe progenitor assumptions, nuclesynthesis, IMF, SNe thermal
energy deposition to the ISM efficiency, etc.  The combination of all these
differences manifests itself in very different predictions for $t_{\rm GW}$.

AY87 and BCF94 represent the extremes in this apparent dichotomy; for massive
ellipticals, the former prefer the late-time galactic wind ($t_{\rm GW}\approx
1$ Gyr), whereas the latter prefer a much earlier time-scale ($t_{\rm GW}\approx
0.1$ Gyr).  On the surface, these results would appear incompatible,
particularly when one considers that star formation is presumed to occur only
for $t\simlt t_{\rm GW}$.  That both groups can claim their final models
match the observations, when one's star formation is an order of magnitude
longer in duration, is puzzling, and it is important to be aware of exactly
how this comes about in order to appreciate the claims of all the groups in
question.

To anticipate what follows, let us look at one massive elliptical scenario:
for an initial gas mass $M_{\rm g}(0)=10^{12}$ M$_\odot$ and Salpeter (1955)
IMF, G96 predict $t_{\rm GW}=0.44$ Gyr, whereas for the same IMF slope, 
BCF94 find $t_{\rm GW}=0.09$ Gyr.  Both the final
colours (\eg V-K$\approx$3.35) and metallicity ($[<{\rm Z}>]_{\rm V}\approx
+0.45$) match the observations.
On the other hand, for a flatter IMF (\eg 
power-law slope, by mass, of $x=0.95$), G96 found $t_{\rm GW}=0.05$ Gyr,
whereas for the same slope, AY87 found $t_{\rm GW}=0.71$
Gyr.  Again, both the colours and metallicities were consistent with those just
mentioned.  
One is left with the question -- \it what are the driving factors which
lead to very different results for $t_{\rm GW}$, and yet still allow each group
to claim reproducibility of the present-day observational constraints? \rm

At some level, this is difficult to answer as there is a tendency to not report
each and every assumption regarding the various input ingredients.  
As such, 
we do not wish to belabour the issue by minutely examining 
every parameter of these codes, but we do feel it important to highlight some
of the primary differences which lead to the, at times, very different results.

\subsection{Gibson (1996b)}
\label{g96}

Our working template, as described in detail in Gibson (1995,1996b),
was generated using the time- and metallicity-independent
Salpeter (1955) IMF, with lower and upper mass limits of
$m_\ell=0.2$ M$_\odot$ and $m_{\rm U}=65.0$ M$_\odot$, respectively.
The metallicity-dependent yields of Woosley \& Weaver (1995) were used for Type
II SNe.  The global
thermal evolution of the ISM was governed by knowledge of the Type Ia and II
SNe rates as a function of time, coupled with the thermal energy made available
to the ISM by each SN, as a function of time.  The model for the
latter assumes that
individual SNRs halt their expansion once
coming into pressure equilibrium with the ambient ISM,
but continue to cool radiatively
\it ad infinitum\rm.  This form is denoted Model B$_2$ in Gibson's (1994b)
notation and is taken directly from Cioffi \etal (1988).
Diffuse dark matter halos with mass and radial extent ratios
relative to the luminous component of ten were used, following the formalism of
Bertin \etal (1992).  3\% of the mass in the IMF in the mass
range $3\rightarrow 16$
M$_\odot$ was assumed to be locked into Type Ia-progenitor binary systems
(\cite{GR83}), an \it a posteriori \rm choice which ensured a present-day Type
Ia SN rate in agreement with Turatto \etal (1994)\footnote{$H_0\equiv 50$
km/s/Mpc assumed throughout.}.  Star formation was assumed
to proceed in lockstep with the available gas mass, the proportionality
constant linking the two, $\nu$, simply being the inverse of the time scale for
star formation.  The chemical evolution is similar in spirit to that of
Matteucci \& Greggio (1986), although we have adopted a more intuitively
obvious ``mass in/mass out'' formalism (Timmes \etal 1995), 
as opposed to the Talbot \& Arnett (1971) ``matrix'' formalism.
The photometric evolution was coupled to the chemical evolution, as outlined in
Gibson (1996a).  In general, the metallicity-dependent isochrones of Worthey
(1994) were adopted, although when those of Bertelli \etal (1994)
were used, the distinction is made.

Table \ref{tbl:templatemodels2} shows some of
the properties of our template of models -- in the first block 
(\cite{S55} IMF), five different initial gas masses (column 1) are shown.
We draw attention to the star formation efficiency
$\nu$ in column 2; for the chosen set of input ingredients, $\nu$ was treated
as a free parameter (as it was in AY87 and BCF94), 
the value shown ensured that the present-day
CML relationships were recovered.  The masses
of gas, oxygen, and iron ejected at time $t_{\rm GW}$ (column 3) are listed in
columns 4, 5, and 6, respectively.  The present-day absolute V-band magnitude
and V-K colour are given in columns 7 and 8.
Column 9 gives the predicted
V-band luminosity-weighted metallicity $[<{\rm Z}>]_{\rm V}$.
The stellar populations of the giant ellipticals in
our template have [Mg/Fe]$_\ast\approx +0.15$, which is only
marginally lower than the observed $\sim +0.2\rightarrow +0.3$ (Worthey \etal
1992).

\begin{deluxetable}{crccccrcr}
\footnotesize
\tablecaption{G96 Template Models\tablenotemark{a}
\label{tbl:templatemodels2}}
\tablewidth{0pt}
\tablehead{
\colhead{$M_{\rm g}(0)$} & \colhead{$\nu\;\;$} & \colhead{$t_{\rm GW}$} & 
\colhead{$m_{\rm g}^{\rm ej}$} & \colhead{$m_{\rm O}^{\rm ej}$} & 
\colhead{$m_{\rm Fe}^{\rm ej}$} & \colhead{M$_{\rm V}$} &
\colhead{V-K} & \colhead{$[<{\rm Z}>]_{\rm V}$}
}
\startdata
\multicolumn{9}{c}{IMF Slope $x=1.35$} \nl
1.0e6  & 188.9 & 0.006 & 3.5e5  & 2.2e2 & 1.0e1 &  -8.21$\quad$ & 2.08 & -2.28$\quad\;$ \nl
5.0e7  & 209.7 & 0.007 & 1.2e7  & 3.2e4 & 1.5e3 & -12.61$\quad$ & 2.12 & -1.56$\quad\;$ \nl
1.0e9  & 123.1 & 0.016 & 1.7e8  & 3.0e6 & 2.0e5 & -15.89$\quad$ & 2.44 & -0.51$\quad\;$ \nl
5.0e10 &  46.0 & 0.077 & 3.4e9  & 1.1e8 & 9.3e6 & -20.15$\quad$ & 3.04 & +0.13$\quad\;$  \nl
1.0e12 &  17.3 & 0.440 & 1.7e10 & 4.9e8 & 7.5e7 & -23.45$\quad$ & 3.33 & +0.44$\quad\;$  \nl
\multicolumn{9}{c}{IMF Slope $x=0.95$} \nl
1.0e12 &  88.3 & 0.049 & 9.6e10 & 6.6e9 & 6.2e8 & -23.03$\quad$ & 3.38 & +0.44$\quad\;$  \nl
\enddata
\tablenotetext{a}{
Input ingredients:  Type Ia and II SNe yields from Thielemann \etal (1993) and
Woosley \& Weaver (1995); low mass stellar yields from Renzini \& Voli (1981).
Photometric properties (columns 7-9) are
based upon Worthey's (1994) isochrones.  SNR energetics
governed by Model B$_2$ of Gibson (1994b).  
The inverse of the star formation timescale $\nu$ [Gyr$^{-1}$] is in column 2.
Column 3 shows the galactic wind time $t_{\rm GW}$ [Gyrs], for the relevant
initial gas masses (column 1 -- M$_\odot$).  The masses of gas, oxygen, and
iron ejected at $t_{\rm GW}$ are listed in columns 7-9 [M$_\odot$].
See Section \ref{g96} for details.}
\end{deluxetable}

The final entry in Table \ref{tbl:templatemodels2} illustrates how choosing the
flatter IMF of slope $x=0.95$ necessitates increasing $\nu$ by a factor of $\sim
5$ in order to ensure that the colours do not become too red, nor the
metallicity too high.  We will
return to this shortly in Section \ref{ay87}.  As an aside, the flatter IMF
leads to an increased value for [Mg/Fe]$_\ast$ of $\sim +0.35$.

In Figure \ref{fig:fig1}, we show the evolution of the elemental abundance
of the primary metals\footnote{For element X,
[X]$\equiv$log X - log X$_\odot$.} for the $M_{\rm
g}(0)=10^{12}$ M$_\odot$ template model of Table \ref{tbl:templatemodels2}.
The evolution shown 
compares favourably\footnote{The abundance ``spike'' immediately following
$t_{\rm GW}$ in Figure \ref{fig:fig1} is not seen in Matteucci \& Padovani's
(1993) Figure 1.  This is simply an artifact of the two codes' temporal
resolutions; in the vicinity of $t_{\rm GW}$, Matteucci \& Padovani use $\Delta
t\approx 50$ Myr, whereas our curves were generated with a uniform $\Delta
t=0.5$ Myr.  The two codes are entirely consistent when identical $\Delta t$
values are assumed.}
with that shown in Figure 1 of Matteucci \& Padovani
(1993).  Recall that
star formation in this model ceases at $t_{\rm GW}=0.44$ Gyr.  
One can see that it only takes $\sim 0.1$ Gyr for any one
of the elements to reach
their approximate solar abundance (within a factor of $\sim 2$).
Note that each of the elements undergoes a
post-$t_{\rm GW}$ dilution due to the continually increasing gas mass (from
dying low mass stars).  The one element which is immune to this dilution is
iron -- the enormous quantity of Fe ejected per SNe Ia event ($\sim 0.74$
M$_\odot$ -- Thielemann \etal 1993) 
is enough to counteract said dilution.  Carbon declines slightly
slower than the other elements (besides iron, of course) after $t_{\rm GW}$ due
to the increased importance of the carbon-producing single low and intermediate
mass stars (\cite{RV81}).  As the Type II yields used for this model 
(\cite{WW95})
did not have any primary nitrogen production, it is not until the $\sim
4\rightarrow 8$ M$_\odot$ stars with primary N (\ie the \cite{RV81}
models with
hot bottom burning included) start dying that [N] starts increasing.  This is
reason why the [N] curve takes $\sim 50$ Myr to approach that of the others.

\placefigure{fig:fig1}

\subsection{Arimoto \& Yoshii (1987)}
\label{ay87}

Recall from Section \ref{comparison} that adopting an IMF slope $x=0.95$, and
adjusting $\nu$ to ensure the proper CML predictions for the $10^{12}$
M$_\odot$ model are recovered, leads to $t_{\rm GW}=0.05$ Gyr, considerably
smaller than the $t_{\rm GW}=0.71$ found by AY87,
yet both claim to replicate the present-day photo-chemical properties for said 
elliptical.  Let us now step through, one-by-one, 
the sources and implications of the discrepancy.  As a
starting point, we use the G96 prediction, given by the final entry to Table
\ref{tbl:templatemodels2}:

\begin{itemize}
\item{\it Star Formation Efficiency: } For the $10^{12}$ M$_\odot$ model,
AY87 required $\nu=8.6$ Gyr$^{-1}$ to recover the proper present-day
photo-chemical properties, whereas we needed $\nu=88.3$ Gyr$^{-1}$.
If we were to reduce $\nu$ by a factor of
ten, our predicted $t_{\rm GW}$ would increase from 0.05 Gyr to 
1.42 Gyr.  The predicted (V-K) and $[<{\rm Z}>]_{\rm V}$,
with this change alone, though, would both be erroneous
(specifically, 4.04 and +0.83, respectively -- recall that V-K$\approx$3.35 and
$[<{\rm Z}>]_{\rm V}\approx +0.45$ were the best values for this luminosity).
\item{\it Type Ia SNe: } AY87 neglect Type Ia SNe entirely.  This immediately
eliminated one potential observational constraint from their modeling (\ie the
magnesium overabundances relative to iron -- Worthey \etal 1992), as this
all-important iron source was not included.  On the other hand,
this is not particularly important for $t_{\rm GW}$;
when using their flat IMF to run the previous model, but this time with Type Ia
SNe ``switched off'', $t_{\rm GW}$ only increases to 1.43 Gyr.
\item{\it Yields: } AY87 use the older Arnett (1978) yields, although the exact
implementation is unclear (in particular, for Z$>$Z$_\odot$).  
Still, replacing the Woosley \& Weaver (1995) yields with those
of Arnett (1991) (which are reasonably similar to his older 1978 yields, as far
as global Z is concerned), led to a mild increase in $t_{\rm GW}$ to 1.51 Gyr.
This comes about because the SNR thermal energy evolution Model B$_2$ is
mildly metallicity-dependent at late-times 
(\ie $\varepsilon_{\rm th}\propto$Z/Z$_\odot^{-0.13}$ -- Cioffi \etal 1988),
and for the same $\nu$, the Arnett (1991) yields lead to a slightly more
enriched gas, in comparison with the Woosley \& Weaver (1995) yields (Gibson
1995,1996b).  The more enriched gas leads to increased cooling, which decreases
the effective energy contribution per SN, and thus increases $t_{\rm GW}$..
\item{\it Dark Matter: } AY87 do not include any dark matter component.
Eliminating this reduces $t_{\rm GW}$ from the previous 1.51 Gyr, to 1.25 Gyr,
due to the slight reduction in the depth of the potential well.
\item{\it ISM Binding Energy: } Ignoring dark matter, AY87 used the gaseous
binding energy formalism of Saito (1979b), as opposed to that of
Bertin \etal (1992), which was used by G96 and BCF94.  
This older form, 
in the absence of dark matter, predicts higher binding energies
for the same residual ISM gas mass.  Using Saito's (1979a) form in lieu of
Bertin \etal's (1992), leads to an increase in $t_{\rm GW}$ from the previous
1.25 Gyr, to 2.06 Gyr.
\item{\it SNe Thermal Energy Evolution: } AY87 use what we (\cite{G94b})
call Model A$_0$, which is simply the classic SNR energetics form due to Cox
(1972) and Chevalier (1974).  On the other hand, G96 use the aforementioned
Model B$_2$, based upon Cioffi \etal (1988).  Switching
to A$_0$ reduces $t_{\rm GW}$ to 1.82 Gyr.  Model A$_0$ neglects any of the
metallicity effects in the cooling of SNRs, so, as noted above,
it is to be expected that there
is a marginal increase in SNR energy efficiency, leading to a slightly
earlier wind (at least for predominantly super-solar metallicity populations,
such as the giant elliptical currently under consideration).  Still, as noted
by Gibson (1994b), Models A$_0$ and B$_2$ are quantitatively similar, despite
their somewhat different approaches, which is why the influence on $t_{\rm GW}$
is not excessive.
\item{\it Stellar Lifetimes: } AY87 use the Talbot \& Arnett (1971)
singular power-law for stellar lifetimes, 
which tends to overestimate, by up to an order of magnitude,
the lifetime of most Type II SNe progenitors.  Using this form for the
lifetimes, instead of the more appropriate Schaller \etal (1992) ones, 
increases $t_{\rm GW}$ again, from 1.82 Gyr to 2.18 Gyr.
\item{\it Hydrogen Number Density - Further SNe Energetics: }
AY87 assumed that the hydrogen number density $n_0$
was constant throughout time, and given by $n_0(t=0)$.  This 
overestimates $n_0$ for $t>0$, which in turn, depletes the available
energy per SN event which is made available to the ISM for powering a galactic
wind, because of the $\sim n_0^{-1/3}$ dependence in the late-time SNR interior
thermal energy evolution (\cite{C74}).
Adopting their invalid assumption results in a large increase in $t_{\rm
GW}$, from 2.18 Gyr, to 10.52 Gyr.  Such a late wind epoch seems highly
unlikely, as, under the current formalism, the implied star formation rates for
ellipticals in the redshift range $z\approx 0.1\rightarrow 0.4$ would be
$\sim 30\rightarrow 40$ M$_\odot$/yr, inconsistent with the observed rates
(\eg \cite{S86}).
This error was also discussed by Angeletti \& Giannone (1990).
\item{\it IMF Limits: } Instead of our template IMF range of $0.2\rightarrow
65.0$ M$_\odot$, AY87 use $0.05\rightarrow 60.0$ M$_\odot$.  This has the
advantage of tying up much more mass in ultra-low mass (\ie long-lived)
objects, thereby
reducing the gaseous binding energy at later times, despite the corresponding
reduction in the absolute Type II SNe production.  Adopting their mass range
reduces $t_{\rm GW}$ from 10.52 Gyr, to 7.81 Gyr.
\item{\it Type II SNe: } To partially
compensate for the lack of Type Ia SNe in their
modelling, AY87
arbitrarily adopt a lower limit for Type II SNe progenitors of 3.0 M$_\odot$.
This is, of course, incorrect on stellar evolution grounds,
but obviously serves as a convenient means to reduce $t_{\rm GW}$
to more reasonable values (as all the $3\rightarrow 8$ M$_\odot$ stars which
would normally end their lives as thermally pulsating asymptotic giant branch
stars would now be assumed to end their lives in a SN explosion) -- 
specifically, $t_{\rm GW}$ is now predicted to be
0.98 Gyr.  At this point, the predicted V-K colour and metallicity $[<{\rm
Z}>]_{\rm V}$ are 3.89 and +0.78, respectively, which are both at odds with the
observed V-K$\approx$3.35 and $[<{\rm Z}>]_{\rm V}\approx +0.45$.
\item{\it Proto-galactic Radius: } In G96, as in most other
galactic wind codes, the radius of the pre-wind proto-galaxy is taken to be
approximately that of the present-day spheroids.  At some level this is
probably incorrect (\eg \cite{H80}; \cite{DS86};
\cite{AG91}; Elbaz \etal 1995), but for lack
of better observational constraints on proto-galactic sizes at the onset of
star formation, this seems the most
conservative approach.  In the published AY87 paper, this also appears
to be the approach taken.
In that case, our previously mentioned $t_{\rm GW}=0.98$ Gyr
should be directly comparable to their value of 0.71 Gyr.  The situation
becomes slightly unclear, though,
when we refer back to their erratum (\cite{AY89}) in which it is
stated that the proto-galactic radius $R_{\rm L}$ used
was actually a factor of two larger than that predicted by the Saito (1979a)
mass-radius relations.  This, in turn, implies that
the proto-galactic binding energy is a factor of two smaller.
Assuming that to be the case, if we re-run the last model, we
find $t_{\rm GW}=0.28$ Gyr.  Equally confusing, 
the preprint version of the AY87 paper states that
they actually adopted a proto-galactic radius which was 
4.74 times that given by Saito (1979a)\footnote{This factor of 4.74, as
quoted in the AY87 preprint, led Matteucci \& Tornamb\`e (1987) to do the
same.  This was corrected for subsequent papers based upon her code
(\cite{M95}).  Angeletti \& Giannone (1990) have likewise commented on this
factor of 4.74.}
Adopting this more extreme initial condition, leads to $t_{\rm GW}=0.03$ Gyr,
again due to the reduction in the depth of the potential well.
The confusing nature of the proto-galactic radius used by AY87 makes
replicating their results, exactly, difficult.
Still, based upon the fact that their published $t_{\rm GW}=0.71$ Gyr, is
closer to what we found when simply using Saito's (1979a) binding energy
prediction ($t_{\rm GW}=0.98$ Gyr), 
as opposed to what was found using arbitrary reductions of
factors of $\sim 2\rightarrow 5$, it would
seem likely that this is what AY87 assumed.
\item{\it Photometric Calibrations: }  AY87 (as described in more detail in
\cite{AY86}) did not have access to any super metal-rich (\ie Z$>$Z$_\odot$)
photometric calibrations).  As such, the best they could do was assume 
that metal-rich dwarfs and
giants obeyed the same colour-luminosity relations as Z=Z$_\odot$ stars.  
For the dwarf
galaxies, whose ISMs do not exceed 
Z$\sim$Z$_\odot$ for most of their star forming
period, this is not a problem.  For the giant elliptical under consideration in
this comparison, though, the bulk of the star formation (\ie $t\simgt 0.05$ Gyr
-- see Table 5 of AY87) occurs while the ISM metallicity
is super-solar, approaching ten
times solar after a few tenths of a Gigayear.  As noted above, 
if we use the full grid of
Worthey (1994) isochrones, including properly the super-solar metallicity ones,
to predict the integrated V-K of the stellar populations at $t=12$ Gyr, we
find V-K$\approx$3.89, which is $\sim 0.55$ mag redder than the mean
observed value for this luminosity.  This would imply that $t_{\rm GW}$ is too
late for the given initial mass/star formation scenario.  On the other hand, if
we impose the condition that all Z$>$Z$_\odot$ stars obey the Z=Z$_\odot$
colour-temperature scale, we would actually predict V-K$\approx$3.28, very much
in line with both the observed mean, and that predicted by AY87.
\item{\it Luminosity-weighted Metallicity: } Instead of calculating the mean
luminosity-weighted metallicity $[<{\rm Z}>]_{\rm V}$, AY87 use the mean of
the
logarithmic, luminosity-weighted metallicity $<[{\rm Z}]>_{\rm V}$.  This is
somewhat
incorrect, as spectral indices scale much closer to Z (\ie number of absorbers)
than to log Z (\eg \cite{W94}; \cite{GG95}).  A simple numerical example
illustrating how this could lead to an underestimation of
the true stellar metallicity is shown in Gibson (1996a).
AY87 do not include any post-red giant branch contribution
(\ie no horizontal branch, asymptotic giant branch, white dwarfs, etc.), which
will play a part at some level, especially the lack of AGB stars, a point which
they themselves acknowledge in their paper.
\end{itemize}

In summary, we feel confident that we now have a good understanding of
the source of the majority of differences in the predictions of AY87 and G96;
of particular importance would appear to be the galactic wind 
time/photometric calibration ``conspiracy''.  

For $M_{\rm g}(0)=10^{12}$ 
M$_\odot$, G96's preferred Salpeter (1955) IMF model predicted 
$t_{\rm GW}=0.44$ Gyr, which is equivalent to $\sim 3\times 10^{48}$ erg/SN 
being made available to the ISM thermal energy reservoir.  In comparison, AY87
favoured a much flatter IMF ($x=0.95$) and $t_{\rm GW}=0.71$ Gyr, which still
corresponds to $\sim 2\times 10^{48}$ erg/SN.  In fact, AY87's
models \it required \rm both a significantly flatter-than-Salpeter (1955) IMF,
and the late galactic wind time, in order to ensure that enough of
the stars formed during the ``super-solar'' metallicity phase
sampled the Z=Z$_\odot$ stellar evolution tracks and their respective
photometric calibrations.   Our models show that AY87's \it necessary \rm
high mass
star biased IMF is no longer a necessity, provided one takes account of the
Z$>$Z$_\odot$ photometric calibrations properly.

On the other hand, if we do impose the $x=0.95$ IMF\footnote{Indeed, such an
IMF is advantageous when attempting to account for the intracluster medium
abundances (\cite{GM96}).}, without altering the star formation efficiency, we
find the final colours/metallicity become too red/rich (V-K=3.97 and $[<{\rm
Z}>]_{\rm V}=+0.77$, respectively), as
$t_{\rm GW}$ does not change dramatically (Gibson 1995,1996b).  
For this flatter IMF, by increasing $\nu$ by a factor of 
$\sim 10$, one can recover the appropriate photo-chemical properties, with
$t_{\rm GW}=0.05$ Gyr.  In this latter scenario, the effective energy
contribution per SN event is found to increase to $\sim 10^{49}$ erg.

In conclusion, AY87 were forced to accept a flatter-than-Salpeter (1955) IMF
because of their lack of metal-rich photometric calibrations.  This would
appear to no longer be a necessity.  Also, at some level AY87 were somewhat
fortuitous in their recovering the present-day photo-chemical properties within
this simple framework, given that the SNR energetics were based upon the $t=0$
ISM density.  Because this would, in general, lead to inordinately late wind
epochs, the lowering of the Type II SN-progenitor lower mass limit to 3
M$_\odot$ was needed in order for the wind epoch to occur at a more feasible
value.  This is very much an example of two wrongs making a right!

\subsection{Bressan, Chiosi \& Fagotto (1994)}
\label{bcf94}

A more recent addition to coupled ``elliptical galaxy 
photo-chemical evolution'' studies, with galactic winds, comes from BCF94.
Their models are similar to AY87, with the primary difference being a
substantial improvement in the input stellar physics.  Specifically,
BCF94 are able to draw upon their impressive ``in-house'' expertise
in stellar evolution theory, generating a self-consistent grid of stellar
tracks from the zero age main sequence to the white dwarf/SN stage,
for a wide range of masses and metallicities (\cite{BBCFN94}).
Super-solar photometric calibrations are provided by the newly-available
Kurucz (1993) model atmospheres (with the odd empirical extension).

BCF94 do not list their yield sources, so one must bear this mind when
comparing any chemical evolution models.  A more serious problem lies in
attempting to replicate their photometric predictions.  It was presumed that
the Bertelli \etal (1994)
metallicity-dependent isochrones were employed ``as is'',
yet comparing the integrated
colours of the simple stellar populations (SSPs) in the BCF94 paper (their
Table 3) with those in the original isochrone paper of Bertelli \etal (1994),
shows that some unspecified
modifications were made for the BCF94 analysis.  This has now
been corroborated by Charlot \etal (1996), and further quantified by Gibson
(1996a).

Let us first just list some of the relevant input ingredients to their models.
Minor differences in G96 and BCF94 can be seen here, and are ignored for the
discussion which follows; we shall only concern ourselves with the most
important ones in the itemized list which follows below:
SNR thermal
energy in the ISM follows the older Model A$_0$ (\cite{C72}; \cite{C74};
\cite{G94b}); a Salpeter (1955)
IMF with $m_\ell=0.1$ M$_\odot$ and $m_{\rm U}=120.0$ is adopted;
the ratio of dark-to-luminous mass and radial extents is taken to be five, and
the Bertin \etal (1992)
formalism followed; the star formation efficiency $\nu$ is assumed to be 20
Gyr$^{-1}$, independent of galactic mass; we shall assume throughout that the
Arnett (1991)
yields were used for the Type II SNe ejecta.\footnote{This assumption is based
solely on the fact that some of the earlier
papers from the Padova Group used yields based upon the related Arnett (1978) 
compilation.}

Recall from Section \ref{comparison} that for $M_{\rm g}(0)=10^{12}$ M$_\odot$,
BCF94 found $t_{\rm GW}=0.09$ Gyr was necessary to recover the proper
present-day photo-chemical properties.  Contrast this with the 0.44 and 0.71
Gyr found, respectively by G96 and AY87.
As we will show now, the comparison which we undertook with BCF94, resulted
in the identification of one or two potential problem areas.  Note that in
places we ignore the small differences in $t_{\rm GW}$ invoked by input
ingredient variations.  Changes at the level shown in the corresponding items of
Section \ref{ay87} are to be expected.

\begin{itemize}
\item{\it SNe Progenitors: } Unlike AY87, BCF94 do include Type Ia SNe, 
but  only in
the calculation of the ISM energetics.  Their role in enriching the ISM (and in
particular the iron abundance) is not considered.  Obviously then, they
are not concerned with predicting [Mg/Fe]$_\ast$ in the resultant stellar
populations, nor with the predicted [$\alpha$/Fe] ratio for the intracluster
medium.
\item{\it Stellar Metallicity Determination: } 
BCF94 do not present luminosity-weighted metallicities, but opt for a
mass-weighted determination.  This is usually the recourse for codes which do
not have a parallel photometric evolution code (\eg \cite{MT87}; \cite{AG90};
\cite{MT94}; Elbaz \etal 1995).  As noted in AY87 and Gibson (1996a), this can,
in some instances, lead one to overestimate the true metallicity of the system,
although this does not impact on the results which follow.
\item{\it Pre-SN Stellar Wind Energetics: } BCF94 have adopted what we feel is
an unrealistic energy formalism for stellar wind energy deposition to the ISM.  
In fact, in the BCF94 models, SNe are a completely inconsequential
component to ISM energetics, and it is the enormous stellar wind energy which
leads to their consistently small values for $t_{\rm GW}$.  This conclusion is
at odds with all the previously mentioned galactic wind studies.
We realize that BCF94 \it needed \rm
to impose an early (\ie $t_{\rm GW}\simlt 0.1$ Gyr)
galactic wind in order to recover the present-day elliptical galaxy CML
relations (see next item below), 
but from a physical standpoint their arguments do not seem sound, as
discussed already in Gibson (1994a)\footnote{Although we readily admit that
stellar wind energetics can play a non-negligible role in powering galactic
winds in dwarf galaxies of mass $M\simlt 10^9$ M$_\odot$ (\cite{G94a}).}.  
Regardless of this physics,
Bressan \etal (1996) and Tantalo \etal (1996) still maintain 
that stellar wind energy of this magnitude is a necessity.

We do not want to repeat the argument of Gibson (1994a) here, but there are two
comments which should be made -- \it First\rm, early galactic wind times of the
sort promoted by BCF94 are entirely feasible with the standard 
\it SNe\rm-driven scenario.  All of the individual SNR models talked about thus
far (\eg Models A$_0$ and B$_2$ of Gibson 1994b)
assume shells expand in isolation \it ad infinitum\rm.  As we saw in previous
sections, this resulted in the effective energy contribution to the ISM
per SN being of the order $10^{48}$ erg.  There is certainly no reason to
expect that such a behaviour is necessarily the correct one.  
As Larson (1974) himself noted, in reality, shells
will come into contact and merge/overlap with neighbouring shells, forming
large superbubbles (\eg \cite{T92}).  Subsequent
SNe will continue to explode inside the low density medium behind the
expanding superbubble, and as such suffer far less from radiative losses.  
This can easily raise the effective energy contribution per SN from $\sim
10^{48}$ erg to $\simgt 10^{49}$ erg.  Larson (1974) claims a value of $\sim
10^{50}$ is actually more appropriate.  This, perhaps, more realistic SNR
evolution model is denoted Model B$_3^\prime$ in Gibson's (1994b) notation.
A template of models (parallel to that shown in Table 
\ref{tbl:templatemodels2}) using such an SNR 
evolutionary scenario  has already been presented
in Gibson \& Matteucci (1996), and will not be discussed further here. 
The \it second\rm, and primary reason for not arguing the pros and cons of the
different SNR evolution models here, is that there appears to be a potential
error in
the BCF94 chemical evolution code, which, when accounted for, 
would seem to remove
the necessity for an early galactic wind in their study.  Let us discuss this
further in the following item, as it is of primary importance to interpreting
the BCF94 results.
\item{\it Chemical Evolution: } 
Figure 9 of BCF94 shows the evolution of the
ISM metallicity
Z$_{\rm g}$, for their favoured suite of models.  With the identical input
parameters, we ran both our code (G96) 
and that belonging to Matteucci (1992) (hereafter, M92),
in order to compare against the BCF94 code, the results of which
are shown in Figure \ref{fig:fig2}.  The top panel shows the evolution of the
system gas mass fraction ($f_{\rm g}\equiv M_{\rm g}(t)/M_{\rm g}(0)$).
Consistency is seen between each of the three codes.  It is the
bottom panel which indicates that something \it may \rm be amiss with BCF94's
chemical evolution code\footnote{We feel reasonably 
confident that the ``problem'' does not lie
in the G96 and M92 codes.  Both use very different approaches to solving the
chemical evolution equations, and were developed entirely independent of one
another.  Further, thanks to the kindness of Leticia Carigi (CIDA) and Frank
Timmes (Chicago), the author was able to compare the chemical evolution of
their codes (\cite{C94} and Timmes \etal 1995, 
respectively) with those of G96 for a
variety of well-defined sample cases.  As for the M92 and G96 curves of Figure
\ref{fig:fig2}, the consistency was excellent.  It seems unlikely that all
four of these chemical evolution codes have made the same error, and one is
left with the impression that the problem lies in the BCF94 code.  It is
difficult to quantify the problem beyond this level without more knowledge
concerning the nature of the BCF94 yield implementation.}.  
If, though,  we were to artificially increase the yields by $\sim
65$\%, then both the G96 and M92 curves would overlay precisely BCF94's.  
How overestimating the yields by $\sim 65$\% would impact upon BCF94's
conclusions proves most interesting.  

\placefigure{fig:fig2}

Figure \ref{fig:fig3} shows the predicted metallicity-luminosity (upper
panel) and colour-luminosity (lower panel) behaviour from two different
observational sources.  The data in the 
upper panel is from Terlevich \etal (1981), and that in 
the lower panel is derived from
Bower \etal (1992).  Overlain on each data-set are several model predictions.

\placefigure{fig:fig3}

Curve 2 of Figure \ref{fig:fig3} shows graphically the predictions of G96's
template (Table \ref{tbl:templatemodels2}).  Recall that the choice of
the star formation efficiency parameter $\nu$ shown in the table
ensured the models agreed with the observations.  Recall, also, that the
galactic wind time $t_{\rm GW}$ was set solely by the SNR energetics (\ie no
pre-SN mass-loss thermalized kinetic energy), and for giant ellipticals,
$t_{\rm GW}\approx 0.4$ Gyr.  

Now, if we assume, as Figure \ref{fig:fig2} suggests, that BCF94 have
overestimated their yields by $\sim 65$\%, we can predict what sort of CML
relation BCF94 \it would \rm have found, \it if \rm they had only used the
conventional Model A$_0$ SNR energetics, with no additional pre-SN stellar wind
energy, along with the incorrect yields.  
For their $\nu=20$ Gyr$^{-1}$ (and Salpeter 1955 IMF with $m_\ell=0.1$
M$_\odot$ and $m_{\rm U}=120.0$ M$_\odot$), said prediction is given by curve 1
in Figure \ref{fig:fig3}, with $t_{\rm GW}\approx 0.4$ Gyr, for the massive
ellipticals, as in G96.  First, the relations are too flat (particularly
in the colour-luminosity plane), but that simply reflects their adoption of a
mass-independent value for $\nu$.  Second, and much more importantly, the
predicted metallicities and V-K colours are $\simgt 0.4$ dex ($\simgt 0.4$
mag) too high (red).  This demonstrates exactly why BCF94 needed to invoke
non-standard stellar wind energetics (recall Gibson 1994a), 
in order
to get the galactic wind to occur early enough so that enrichment could not
drive the relations too rich/red.

Curve 4 resembles curve 1 in that the Arnett (1991) yields have been scaled
upwards by $\sim 65$\%, \it but \rm we have imposed the galactic wind times,
as given by BCF94, to each of the models.  These should be the direct analog of
BCF94's models.  These latter models are denoted as curve 3 in Figure
\ref{fig:fig3}.  The metallicity-luminosity agreement between the two is
excellent (especially when we consider that BCF94 used a mass-weighted
metallicity determination in lieu of a luminosity-weighted one).  The
(V-K)--M$_{\rm V}$ relation given by
curve 4 is offset by $\sim 0.08$ mag from curve 3,
because of the redder giant branch in Worthey's (1994)
Z=Z$_\odot$ isochrones.  This well-known effect has
already been discussed by Gibson (1996a) and Charlot \etal (1996).
One can now see, especially from the lower panel of Figure \ref{fig:fig3},
why BCF94 preferred their early wind formalism, as it gives the appearance of
solving the problem presented by curve 1.  We presume that BCF94 were
only concerned with recovering the colour-luminosity relation in the bottom
panel, as the metallicity-luminosity relation of the upper panel for their
curve 3 (or curve 4, for that matter), lies below the Terlevich \etal (1981)
calibration.  On the other hand, the Terlevich \etal (1981) calibration may
overestimate the true metallicity by $\sim 0.1\rightarrow 0.2$ dex
(\cite{B94}), so we should not overinterpret this particular constraint.

On the other hand, what might we expect if BCF94's early wind epochs (\ie
dominant pre-SN stellar wind energy) were appropriate, \it and \rm the yields
were brought back down in line with Arnett's (1991) tables?  The answer is
provided by curve 5 in Figure \ref{fig:fig3}.  For the given IMF and star
formation efficiency, \it if \rm the yields were accounted for properly, BCF94
would have found metallicities (colours) which were 
too low (blue) by $\sim 0.5$ dex ($\sim 0.2$ mag).  We confirmed this with
Matteucci's (1995) mass-weighted metallicity determination code, as applied to
models derived from her 1992 chemical evolution
code, with enforced wind times to
match those of BCF94.  Curve 6 of Figure \ref{fig:fig3} illustrates this
result (which parallels that of curve 5, as it should).
\end{itemize}

In summary, BCF94, Bressan \etal (1996), and Tantalo \etal (1996) have each
stressed that the primary motivation for invoking early galactic winds 
(\ie $t_{\rm GW}\simlt 0.1$ Gyr) via thermalized kinetic energy from
pre-SN mass-loss, was not necessarily a ``physical'' one, but more out of
necessity, in order to recover the present-day CML relations.  As we have
demonstrated above, what drove them to this conclusion appears to have been an
overestimation of the metal yields (by $\sim 65$\%) in their chemical evolution
code.  Correcting this apparent problem removes their entire motivation for
looking for an energy source capable of powering early galactic winds in the
first place!  

This is further corroborated by the downward ``extensions'' to
curve 1 of Figure \ref{fig:fig3}.  
Here the $10^{12}$ M$_\odot$ models are
shown by the asterisks, along curve 1 in both panels,
where BCF94 would have predicted it to be, assuming no stellar wind energy and
Arnett (1991) yields scaled upwards by $\sim 65$\%. Again, the $t_{\rm GW}$ is
$\sim 0.4$ Gyr, in this case.  \it If \rm BCF94 had simply retained the SN
energetics, but used the Arnett (1991) yields, as is, the predictions would
shift to the lower part of the respective extensions.  As one can see, the
values are in-line with the observations, and any further ``tweaking'' could be
accomplished by modifying $\nu$ upwards, slightly.  Once again, the point to be
made is that ``late'' galactic winds (\ie $t_{\rm GW}\approx 0.4$ Gyr) are more
than adequate to recover the CML relations, \it provided \rm the chemical
evolution properly reflects the SNe yield predictions.

\section{Summary}
\label{summary}

A comparison of the galactic wind codes of AY87 and G96 demonstrates quite
clearly that AY87's conclusion that the IMF in elliptical galaxies \it must \rm
be flatter-than-Salpeter (1955), with a slope, by mass, of $x=0.95$, is purely
an artifact of missing metal-rich (Z$>$Z$_\odot$) photometric
calibrations.  Two erroneous assumptions in their model (that the ISM density is
not a function of time, and that all stars of mass $m\ge 3$ M$_\odot$ end their
lives as Type II SNe) fortuitously combine to yield wind times and
photo-chemical properties which resemble the present-day observations.

In referring to the early wind predictions of BCF94, it would appear that
the criticism of Gibson (1994a) was incomplete.  At that point we had argued
against pre-SN mass-loss via stellar winds playing such a prominent role in
setting $t_{\rm GW}$, purely from a physical standpoint, \it but \rm
that one could
still get early winds via proper handling of conventional SN energetics (see
also G96 and \cite{GM96}).   We still stand by these arguments, but, more
importantly (and a point which we did not appreciate in the earlier
\cite{G94a}), as we have demonstrated in Section \ref{bcf94},
it would appear that BCF94's argument for an early wind, in the first place,
was predicated upon a substantially different, and perhaps incorrect,
treatment of nucleosynthetic yields in their
chemical evolution code.  Scaling BCF94's yields down to the tabulated Arnett
(1991) values removes the \it necessity \rm for the early wind.  It \it may \rm
still occur early (\ie $t_{\rm GW}\simlt 0.1$ Gyr) in the evolution (\eg
\cite{GM96}), but it is no longer a \it necessity\rm.

More sophisticated models, than those discussed here, are needed before the 
``early'' versus ``late'' wind-dichotomy can be
resolved.  Specifically, do SNe contribute $\simlt 10^{48}$ ergs (late) or 
$\simgt 10^{49}$ ergs 
(early)?  Even knowledge at this level may not be enough to
resolve the problem, as it also depends intimately upon the assumed IMF and
star formation formalism.  For the simple models discussed here, with star
formation proportional to the available gas mass, all we can really say is that
a standard Salpeter (IMF) with late (\ie $t_{\rm GW}\simgt 0.4$ Gyr for giant
ellipticals) winds successfully recovers the present-day CML relations, \it
but \rm then so do early (\ie $t_{\rm GW}\simlt 0.05$ Gyr) wind models with a
flatter-than-Salpeter (1955) IMF of slope $x\approx 1$.  One can imagine that
other combinations are potentially available.
At some level, the arguments of Mihara \& Takahara (1994) for simply
treating $t_{\rm GW}$ as a free parameter seem appealing, although we still feel
it important to at least be aware of the different claims from the groups
discussed in this paper, and understand how and why the differences come about.

\acknowledgments

I wish to thank Francesca Matteucci for her ongoing support.  Confirming the
stability of my own chemical evolution 
code has been made infinitely easier by the willingness of
her, Frank Timmes, and Leticia Carigi to run numerous test cases for me, with
their own code, with
which to compare.  The financial support of NSERC, through its Postdoctoral
Fellowship program, is gratefully acknowledged.


%
%

\clearpage

\figcaption[fig1.eps]{Time-dependence of the elemental abundances
(relative to solar) for the $M_{\rm g}(0)=10^{12}$
M$_\odot$, SNe energy Model B$_2$, template of G96.  
The galactic wind epoch $t_{\rm
GW}$ (and hence point of cessation of star formation) is noted.
\label{fig:fig1}}

\figcaption[fig2.eps]{Evolution of the gas mass fraction $f_{\rm
g}$ (upper panel) and global metallicity Z$_{\rm g}$ for 
the $\nu=20$ Gyr$^{-1}$ models of G96, 
M92's chemical evolution code, and BCF94.
The Arnett (1991) Type II SNe yields were adopted for the G96
curves.  The contribution from SNe Ia is neglected.
\label{fig:fig2}}

\figcaption[fig3.eps]{\it Upper Panel: \rm Mean stellar metallicity versus 
absolute V magnitude.  Observational data from Terlevich \etal (1981).
Curve 1: Projected BCF94 behaviour assuming no stellar wind energy 
contribution, and 1.64$\times$ the standard Arnett (1991) yields.  Downward
extension illustrates the anticipated
shift to lower metallicities had BCF94 simply used the tabulated Arnett 
(1991)
yields.  Curve 2: G96 behaviour given by Table \ref{tbl:templatemodels2}.
Curve 3: BCF94's quoted behaviour, from their Table 5.  Curve 4: As curve 1,
except with stellar wind energy of the form used in their paper included.
Curve 5: As curve 4, except tabulated Arnett (1991) yields adopted.  
Curve 6:
As curve 5, except using mass-weighted metallicity code of Matteucci 
(1995), as applied to models derived from Matteucci (1992).
\it Lower Panel: \rm Optical-Infrared colour versus absolute V magnitude.
Observational data from Bower \etal (1992). 
\label{fig:fig3}}

\end{document}